\documentclass[sigconf, language=french,language=german, language=spanish, language=english]{acmart}

\AtBeginDocument{%
  \providecommand\BibTeX{{%
    \normalfont B\kern-0.5em{\scshape i\kern-0.25em b}\kern-0.8em\TeX}}}
\setcopyright{none}
\usepackage{tikz}
\usepackage{multirow}

\usetikzlibrary{shapes,arrows,positioning}

\usepackage{amsmath}
\usepackage{qcircuit}

\renewcommand\footnotetextcopyrightpermission[1]{} % removes footnote with conference
\begin{document}

\title{Towards Advantages of Parameterized Quantum Pulses}
% \translatedtitle{french}{Le nom du titre est l'espoir}
% \translatedtitle{german}{Der Name des Titels ist Hoffnung}
% \translatedtitle{spanish}{El nombre del título es esperanza}

\author{Zhiding Liang}
\email{zliang5@nd.edu}
\affiliation{%
  \institution{University of Notre Dame}
  \city{Notre Dame}
  \state{IN}
  \country{USA}
}
\author{Jinglei Cheng}
\email{cheng636@purdue.edu}
\affiliation{%
  \institution{Purdue University}
  \city{West Lafayette}
  \state{IN}
  \country{USA}
}

\author{Zhixin Song}
\email{zsong300@gatech.edu}
\affiliation{%
  \institution{Georgia Institute of Technology}
  \city{Atlanta}
  \state{GA}
  \country{USA}}

\author{Hang Ren}
\email{hangren@berkeley.edu}
\affiliation{%
  \institution{University of California, Berkeley}
  \city{Berkeley}
  \state{CA}
  \country{USA}}
  
\author{Rui Yang}
\email{ypyangrui@pku.edu.cn}
\affiliation{%
 \institution{Peking University}
 \city{Beijing}
 \country{China}}

\author{Kecheng Liu}
\email{2100013065@stu.pku.edu.cn}
\affiliation{%
 \institution{Peking University}
 \city{Beijing}
 \country{China}}
 
\author{Peter Kogge}
\email{kogge@nd.edu}
\affiliation{%
  \institution{University of Notre Dame}
  \city{Notre Dame}
  \state{IN}
  \country{USA}
}

\author{Tongyang Li}
\email{tongyangli@pku.edu.cn}
\affiliation{%
 \institution{Peking University}
 \city{Beijing}
 \country{China}}

\author{Yongshan Ding}
\email{yongshan.ding@yale.edu}
\affiliation{%
  \institution{Yale University}
  \city{New Haven}
  \country{USA}}

\author{Yiyu Shi}
\email{yshi4@nd.edu}
\affiliation{%
  \institution{University of Notre Dame}
  \city{Notre Dame}
  \state{IN}
  \country{USA}
}

% %%
% %% By default, the full list of authors will be used in the page
% %% headers. Often, this list is too long, and will overlap
% %% other information printed in the page headers. This command allows
% %% the author to define a more concise list
% %% of authors' names for this purpose.


\begin{abstract}
The advantages of quantum pulses over quantum gates have attracted increasing attention from researchers. Quantum pulses offer benefits such as flexibility, high fidelity, scalability, and real-time tuning.
However, while there are established workflows and processes to evaluate the performance of quantum gates, there has been limited research on profiling parameterized pulses and providing guidance for pulse circuit design.
To address this gap, our study proposes a set of design spaces for parameterized pulses, evaluating these pulses based on metrics such as expressivity, entanglement capability, and effective parameter dimension.
Using these design spaces, we demonstrate the advantages of parameterized pulses over gate circuits in the aspect of duration and performance at the same time thus enabling high-performance quantum computing.
Our proposed design space for parameterized pulse circuits has shown promising results in quantum chemistry benchmarks.
\end{abstract}

\maketitle
\pagestyle{plain}
\section{Introduction}

In recent years, significant advancements have been made in both hardware and software development in the field of quantum computing. Quantum computing is a highly promising emerging technology with the potential to solve complex problems that classical computers cannot handle~\cite{Preskill2018NISQ}. Among various technologies used to implement quantum computers, superconducting quantum computers have shown promising results in various applications, such as simulating quantum systems~\cite{arute2020hartree}, optimization~\cite{Google2021QAOA}, and machine learning~\cite{huang2021power}. However, the technology is still in its early stages and faces significant challenges such as decoherence, noise, and scalability~\cite{Preskill2018NISQ}, which must be overcome before it can become a practical tool for solving real-world problems.

In the Noisy Intermediate-Scale Quantum (NISQ) era, superconducting qubits have emerged as the leading technology for developing quantum computers. Superconducting quantum hardware offers several advantages over other types of quantum computing technologies, including scalability, fast gate operations, and flexibility~\cite{cheng2023noisy}. Therefore, we choose superconducting quantum computers as our platform for evaluating parameterized quantum pulses. Although our techniques can be generally applied to other types of quantum computers, such as trapped ion, we focus on superconducting systems due to the constraints related to pulse-level access.

\begin{figure}[t]
\centering
\includegraphics[width=\linewidth]{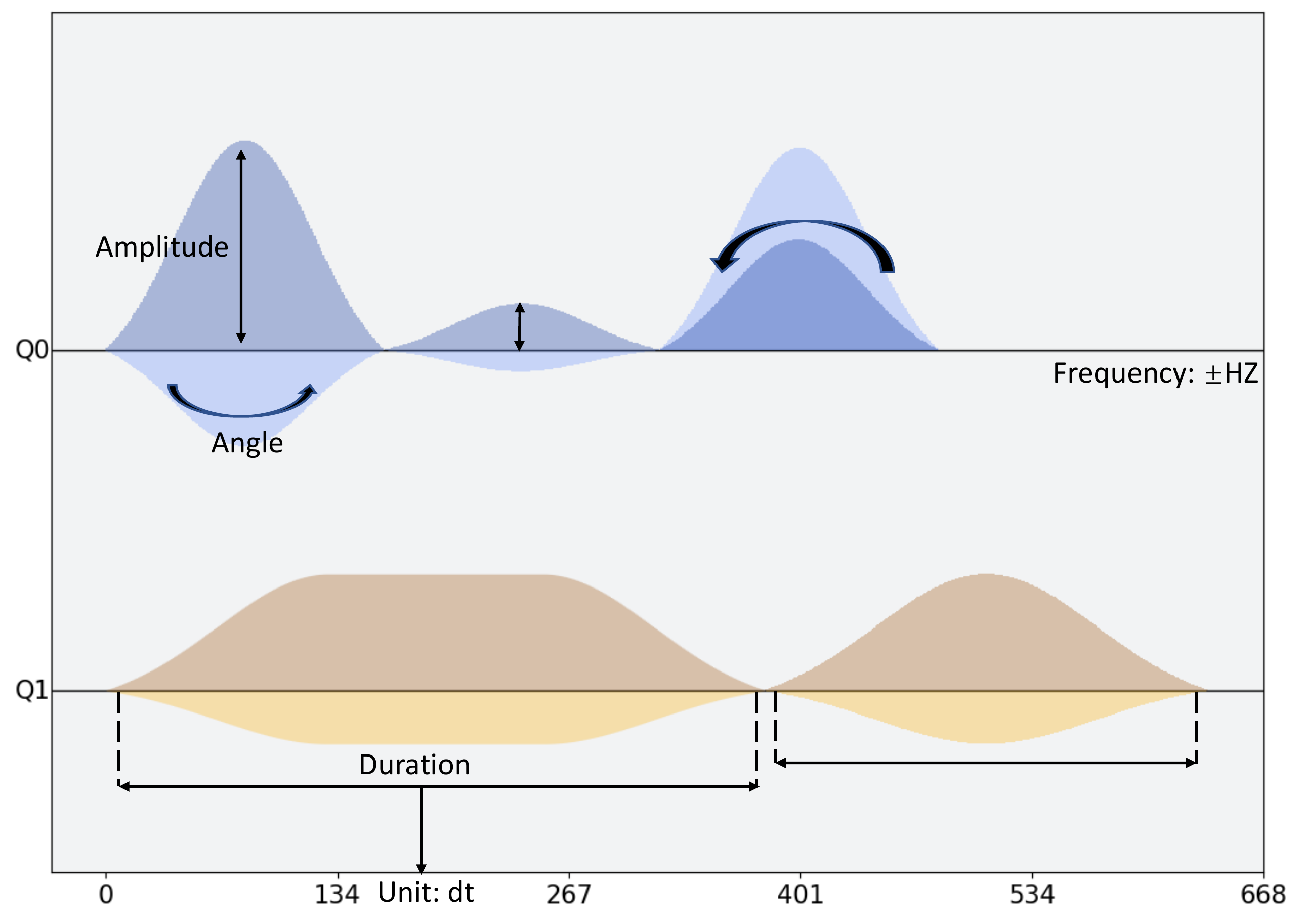}
\caption{
Illustrations of parameterized pulse circuits. Parameters include amplitude, angle, duration, and frequency. The amplitude of a quantum pulse refers to the strength of the signal that is applied to the qubit. The duration refers to the length of time that the pulse is applied to the qubit. The angle determines the phase of the pulse. The frequency specifies the frequency of the carrier signal.
}

\vspace{-4mm}
\label{teaser}
\end{figure}

Quantum programs are usually written in high-level programming languages, such as Q\#, Python, or Quil~\cite{svore2018q,johansson2012qutip,smith2016practical}, which are designed to avoid low-level details about the physical implementation of quantum computing. 
These languages typically provide constructions of quantum circuits, gates, measurements, and other operations.
Once the quantum program is written in a high-level language, it needs to be optimized and transformed into a format that can be efficiently executed on real quantum hardware. 
This involves techniques such as gate fusion, gate cancellation, and gate commutation, which are used to reduce the number of gates and improve the circuit's efficiency.
The next step is to map the optimized circuit onto the physical qubits of the quantum hardware. 
This involves assigning each logical qubit in the circuit to a physical qubit on the hardware, taking into account factors such as connectivity constraints and the availability of resources such as gates and measurements. 
The overall goal is to make quantum circuits compatible with hardware topology while minimizing the total number of SWAP gates that needs to be inserted.
Next, these mapped circuits need to be decomposed into basis gates that are natively supported by the quantum backend.
At the current phase, the compilation remains at the gate level. 
Pulse-level controls come into play when we need to interact with physical qubits.
For superconducting quantum computers, once the decomposition stage is finished, the circuits that are composed of basis gates will be ``translated'' into pulses. 
Finally, these pulses will be transmitted to physical qubits as control signals.

\begin{figure}[t]
\centering
\includegraphics[width=\linewidth]{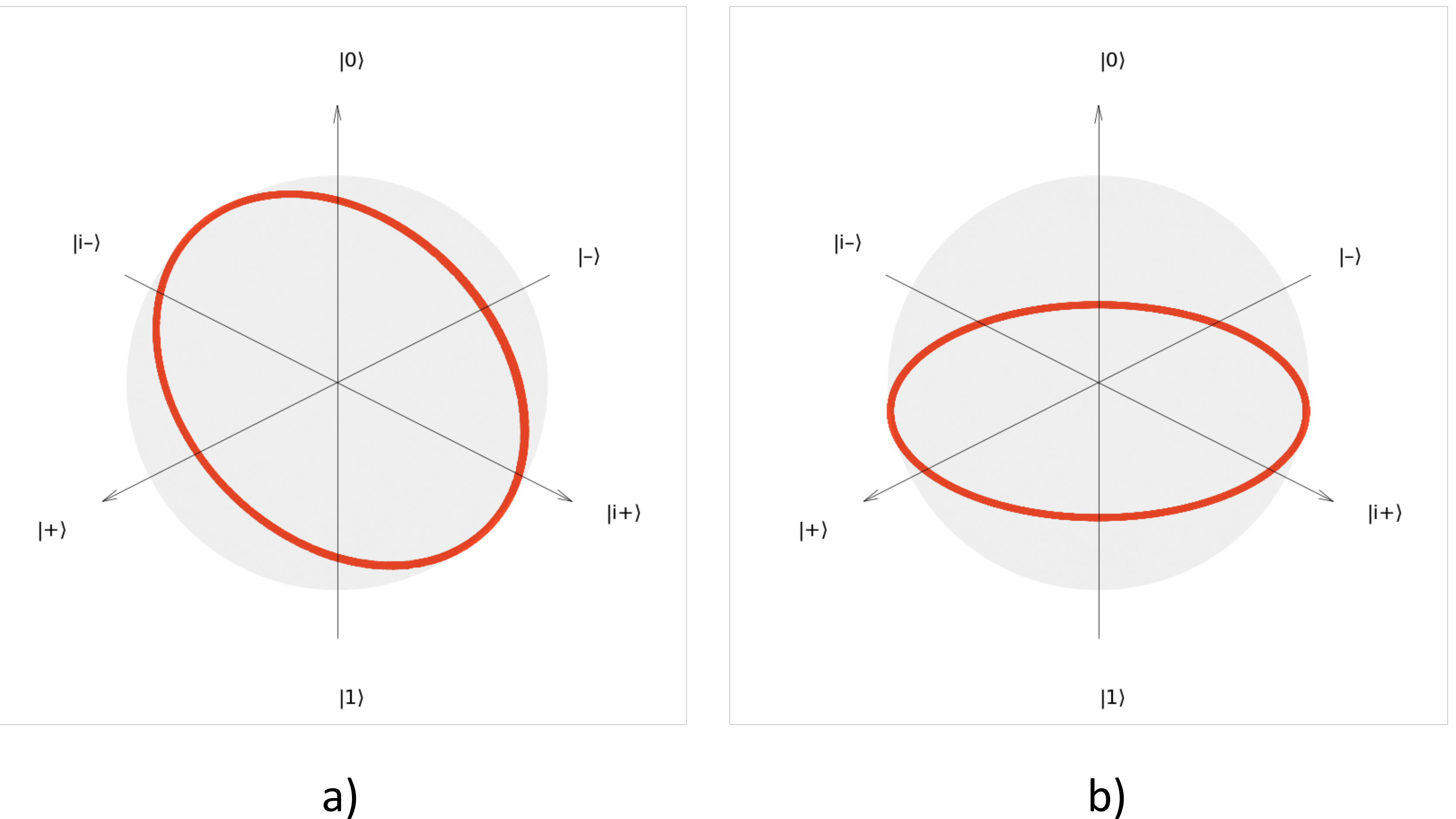}
\caption{
Illustrations of the function of parameterized single qubit pulse. a) Samplings 5000 times for amplitude while fix angle as 0. b) Samplings 5000 times for angle while fix amplitude = 0.08. The operation of amplitude indicates the rotation with respect to the X-axis and the change of angle is corresponding to the rotation with respect to the Z-axis.}

\vspace{-4mm}
\label{oneqpulse}
\end{figure}

Recent studies~\cite{melo2022pulse, egger2023study,peng2023simuq,koch2022quantum,di2023quantum, gokhale2020optimized,shi2019optimized,earnest2021pulse,li2022pulse,smith2022programming} have highlighted the benefits of utilizing pulse-level controls over gate-level programming for a specific set of quantum algorithms, particularly variational quantum algorithms. The key insight is that the gate-level abstraction layer is designed with precise calibration of basis gates for higher fidelities. The calibration of these basis gates aims to minimize their deviation from the ideal theoretical representation. However, the ansatz circuits in variational quantum algorithms may not demand such accurate implementation. 
The "training" process of variational quantum algorithms can inherently correct such imprecision. By incorporating parameterized pulses within ansatz circuits, the search for the desired state in the Hilbert space can be achieved with much shorter circuit latency in comparison to gate circuits. For instance, the current implementation of a U3 gate on IBM's quantum hardware would require two well-calibrated pulses and one virtual-Z operation for execution, whereas at the pulse level, a single pulse can achieve the same functionality.

At present, the majority of quantum programs are designed and executed at the gate level, with a well-established workflow in place for profiling the performance of quantum gates on quantum hardware~\cite{aleksandrowicz2019qiskit, preskill2018quantum}.
Profiling the performance of quantum gates involves measuring and analyzing the physical characteristics and behavior of the gates used in a quantum circuit. 
This information can be leveraged to understand the limitations and sources of errors in the gates, optimize their performance, and improve the overall fidelity and accuracy of the quantum computation.
By understanding the performance of individual quantum gates, researchers can optimize the design of quantum circuits to minimize error rates and improve the overall performance of quantum algorithms~\cite{wang2022quantumnas,wang2022quantumnat,wang2022qoc,wang2021exploration,hu2022quantum,jiang2021co,cheng2022topgen,wang2022quest,qi2023theoretical, pistoia2021quantum,li2021sublinear,wang2021variational,patel2022charter,zhang2022uniq}.
Profiling the performance of quantum gates also enables benchmarking of different quantum hardware platforms~\cite{tomesh2022supermarq,lubinski2021application,larocca2022diagnosing}, and it is crucial for developing error correction and mitigation techniques~\cite{das2021adapt,hua2021autobraid,reed2012realization}.
In summary, profiling the performance of quantum gates is vital for improving the overall performance and reliability of quantum computing systems.
Consequently, profiling parameterized pulses is crucial for the advancement of pulse-level controls and pulse-level quantum programming. We employ various criteria as metrics for evaluating parameterized pulses and examine different pulse templates to investigate their differences. Our aim is to offer guidance for the design and application of parameterized pulses in quantum circuits.

We attempt to build detailed guidance surrounding the topics related to parameterized quantum pulses. To this end, our contribution includes:
\begin{itemize}
    \item A set of pulse-level design spaces provided that guides the design of parameterized quantum pulses of good quality.
    \item A group of criteria to profile the pulse-level design space and characterize the property and power of parameterized quantum pulses.
    \item Benchmark applications include quantum chemistry tasks and quantum finance tasks to validate the effectiveness of proposed pulse-level design spaces.
\end{itemize}

The rest of the paper is organized as follows. We provide a concise overview of parameterized quantum circuits and systems, followed by a discussion on related work in Section 2.
In Section 3, we outline various criteria for pulse-level circuit design and demonstrate the results on benchmarks.
In Section 4, we propose the pulse-level design space which we examine with the aforementioned criteria. 
Then, we employ the parameterized pulses in different applications and obtain the performance results in Section 5. 
In Section 6, we discuss the advantages and limitations of parameterized quantum pulses and present an outlook for future work.

%%%%%%%%%%%%%%%%%%%%%%%%

\begin{figure*}[t]
\centering
\includegraphics[width=\linewidth]{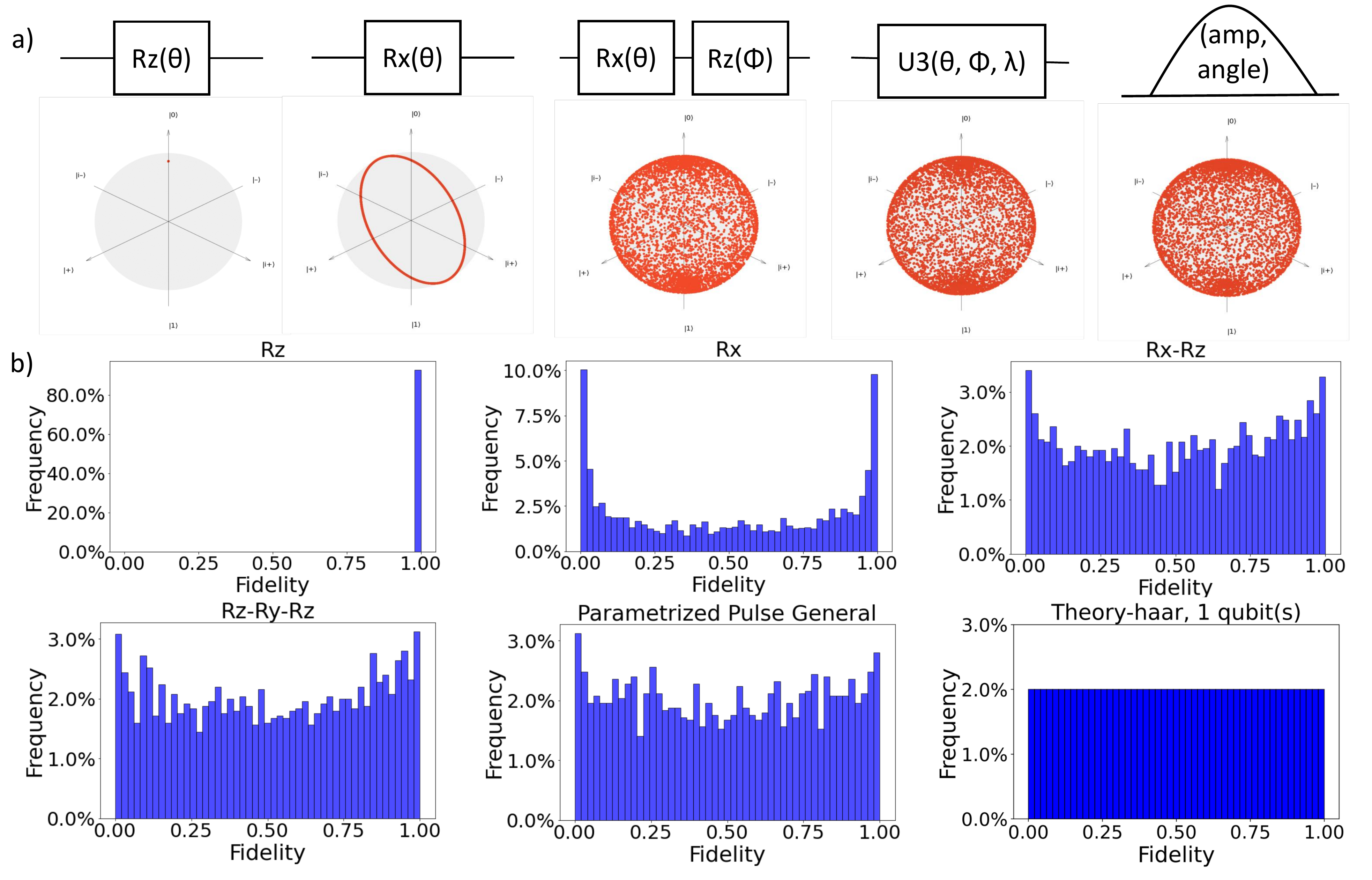}
\caption{
Comparison of the expressivity of single qubit gate-level circuit and single qubit pulse circuit. a) Four different single qubit gate-level circuits: Rz gate, Rx gate, Rx + Rz circuit, and U3 gate (Rx + Rz + Rx), with sequence duration 0dt, 320dt, 320dt, and 320dt, respectively. And a single qubit pulse circuit that takes amplitude and angle as parameters, with a sequence duration of 160dt. For each circuit, we use Qiskit-Dynamics for 5000 times sampling on Bloch Sphere. b) The histograms showcase the estimated fidelities of the enumerated circuits, with a theoretical Haar-distributed fidelity plotted as a reference.}

\vspace{-2mm}
\label{oneqpulseexpr}
\end{figure*}
\section{Parameterized Pulse Circuits}
The investigation of pulse-level quantum computing has gained significant attention from researchers. In this study, we focus primarily on parameterized quantum pulses. 
Most prior applications of parameterized quantum pulses have been focused on the calibration of a quantum computer's basis gates.
For example, experiments on Rabi oscillations are carried out to calibrate the angle and amplitude of single-qubit gates~\cite{sheldon2016characterizing}. 
To calibrate two-qubit gates, Hamiltonian tomography is employed, with the aim of maximizing the proportion of ZX interaction, as it is a crucial element of the CNOT gate~\cite{sheldon2016procedure, mckay2018qiskit}. 
Several runtime-efficient and noise-resilient strategies have been proposed for parameterized quantum pulses, allowing for precise calibration of multi-qubit gates~\cite{mckay2017efficient,sheldon2016procedure, danin2023procedure}. 
Quantum optimal control (QOC) is regarded as one of the most promising options for generating parameterized pulses for a given unitary matrix. Nevertheless, QOC is excessively costly~\cite{khaneja2005optimal, caneva2011chopped,sivak2021model, cheng2020accqoc,giannelli2022tutorial}. 

As depicted in Fig.~\ref{teaser}, typical pulse-level parameters for a superconducting quantum computer consist of amplitude, angle, duration, and frequency. Adjusting these parameters affects the driving Hamiltonian of the quantum operation, thereby impacting the state of the quantum circuit~\cite{magann2021pulses}. Given an example to understand the parameters function in the single qubit parameterized pulse as follows \cite{krantz2019quantum}:
\begin{equation}
\widetilde{H}_d = -\frac{\Omega}{2} V_{0}s(t) (I\sigma_x + Q\sigma_y)
\end{equation}
where $s(t)$ is a dimensionless envelope function, so that the amplitude of the drive is set by $V_{0s}(t)$. Thus, we know amplitude determines the rotation speed, while the duration determines the rotation time. For the function of the angle, refer to the definition of I and Q \cite{krantz2019quantum}:
\begin{align}
\vspace{-1mm}
I = \cos(\varphi) \quad(the\ ‘in-phase’\ component)     \\
Q = \sin(\varphi) \quad(the\ ‘out-of-phase’\ component)
\vspace{-6mm}
\end{align}
$I$ is the in-phase pulse which corresponds to rotations around the $x$-axis, while $Q$ is the out-of-phase pulse which corresponds to rotations about the $y$-axis. $I$ and $Q$ are both dependent on parameter angle $\varphi$, it can be inferred that the angle determines the position of the rotation axis in the XY plane.

Prior research has investigated the optimization of pulse amplitude, frequency, and duration to mitigate decoherence and attain superior performance in calculating the ground state energy of chemical molecules~\cite{meitei2021gate, liang2022pan, egger2023study}, in classification tasks of machine learning~\cite{liang2022variational,pan2023experimental,meirom2022pansatz}, and in the max-cut problem of QAOA on different quantum hardware including superconducting quantum computers~\cite{liang2022hybrid, ibrahim2022pulse} and neutral atom quantum computers~\cite{de2023pulse}.

\section{Pulse-level Circuit Criteria}

To measure the efficacy of pulse-level circuits and emphasize their superiority over gate-level circuits, we present four metrics: expressivity, entangling capability, effective parameter dimension, and sequence duration.
Previous research on parameterized quantum circuits (PQC) at the gate level has utilized both expressivity and entangling capability~\cite{sim2019expressibility,dur2001entanglement}, where sampling random circuits are used as benchmarks. 
In addition, effective parameter dimension is a criterion for analyzing redundant parameters in gate-level quantum circuits. Nonetheless, our objective is to investigate parameterized quantum circuits at the pulse level, namely parameterized pulse circuits (PPC). To evaluate the effectiveness of parameterized pulse circuits, we compare their deviation with that of random circuits, as well as the deviation of PQC with random circuits. By shifting from gate-level to pulse-level, we can achieve improvements in sequence duration while maintaining similar expressivity.

\subsection{Sequence Duration}
The term sequence duration refers to the total time required for the execution of a quantum program.
Sequence duration plays a vital role in determining the overall execution time of the program and can significantly impact its performance. Given that the current decoherence time on quantum machines is limited, sequence duration is a critical consideration when designing quantum programs. Thus, optimizing the sequence duration involves adjusting the pulse parameters, adding delays or idle pulses, or utilizing different pulse shapes to minimize the overall duration and enhance program performance.

In our Qiskit-Pulses test cases, the sequence duration of a pulse schedule is always defined using a basic unit dt, where dt is the duration of a single sample of the arbitrary waveform generator, typically equal to 0.222 ns. 
This duration is discretely mapped from the hardware level to the software level using the unit dt. 
While the search space of duration and amplitude partially overlap, they jointly determine the ``strength'' of pulses. 
As hardware is usually more precise for amplitude control, the hardware control typically selects a fixed duration and then performs amplitude control. 
We decide to retain duration control because we aim to adaptively reduce the decoherence time of the quantum algorithm.
\begin{table}[h!]
\centering

  \captionsetup{}
  \setlength{\tabcolsep}{1.3em}
    \begin{tabular}{|l|c|c|c|}
    \hline
     \textbf{Expressivity} & \textbf{Quito} & \textbf{Lima} & \textbf{Jakarta} \\
    \hline
     expr\_z     &  3.8918   &    3.8918    &      3.8918    \\
     expr\_x     &  0.2344   &    0.1771    &      0.1752   \\
     expr\_xz    &  0.0222   &    0.0202    &      0.0177   \\
     expr\_zyz   &  0.0173   &    \textbf{0.0169}    &      0.0235   \\
     expr\_amp   &  0.1985   &    0.1913    &      0.1673    \\
     expr\_ang   &  0.1709   &    0.7790    &      0.1719    \\
     expr\_pulse &  \textbf{0.0138}   &    0.0202    &   \textbf{0.0157}     \\
     \hline
  \end{tabular}
\caption{The expressivity of different single qubit circuits sampled on three different backend models by Qiskit-Dynamics, expr\_z, expr\_x, expr\_xz,  expr\_zyz, expr\_amp, expr\_ang, expr\_pulse correspond to Rz gate, Rx gate, Rx + Rz circuit, Rx + Ry + Rz (U3 gate) circuit, single qubit pulse circuit with amplitude sampled while fix angle as 0, single qubit pulse circuit with angle sampled while fix amp as 0.08, general single qubit pulse circuit with angle and amplitude all sampled.}
\label{expr}
\vspace{-8mm}
\end{table}

\begin{figure}[t]
\centering
\includegraphics[width=\linewidth]{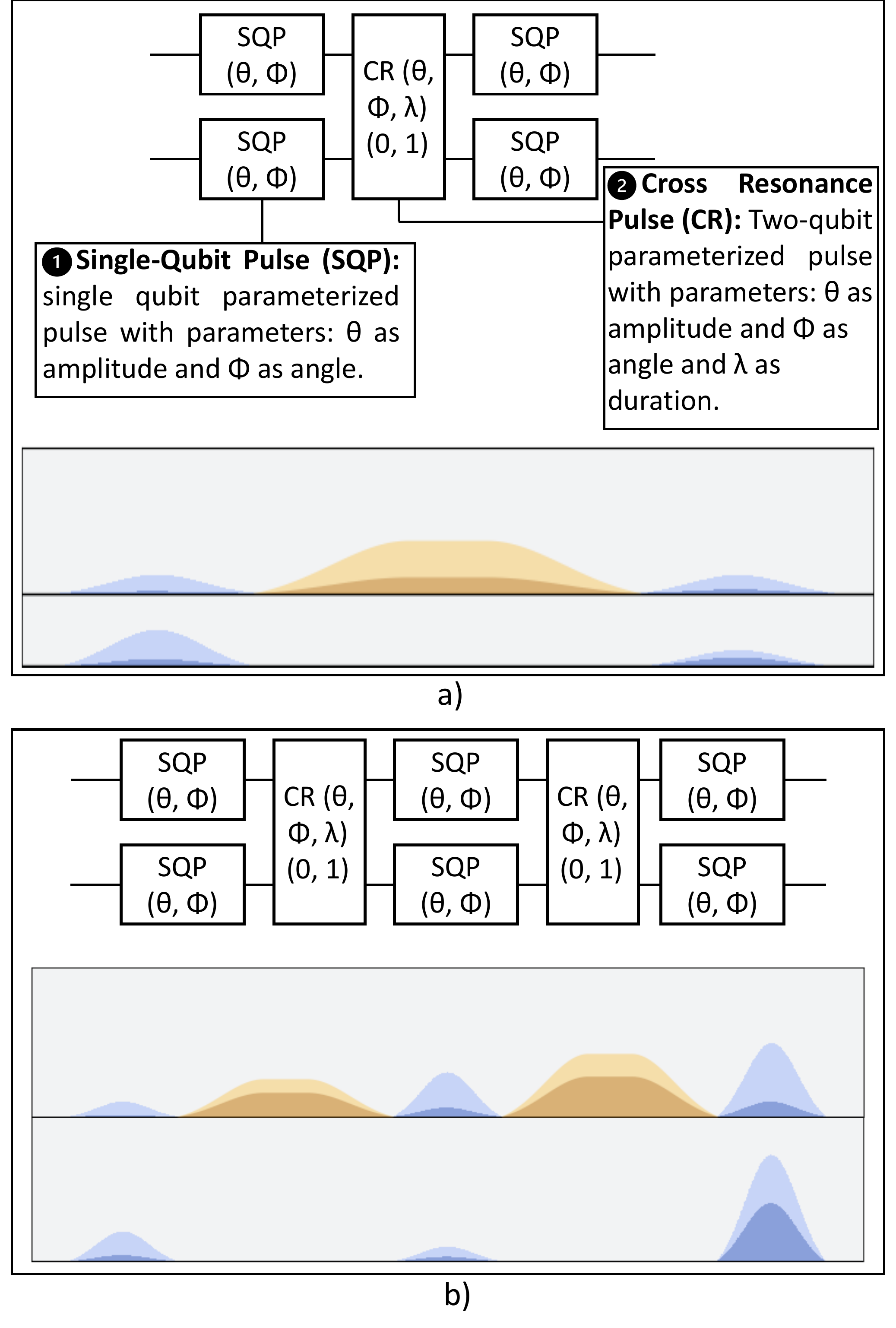}
\caption{Two-qubit pulse circuit structure and corresponding pulse schedule. a) Dressedpulse consists of four single-qubit pulses (SQP) and one cross-resonance pulse (CR). b) Blockpulse consists of six SQPs and two CRs.}

\vspace{-4mm}
\label{2qpulse}
\end{figure}

\subsection{Expressivity}

\begin{table*}[h!]
\centering

  \captionsetup{}
  \setlength{\tabcolsep}{2.5em}
    \begin{tabular}{|l|c|c|c|c|c|}
    
    \hline
     \textbf{Model} & \textbf{Expr} & \textbf{Ent} & \textbf{Param} & \textbf{EPD} & \textbf{Dur}\\
    \hline
     2QDressedpulse  &  0.0259  &   0.1145   & 11&11  & 276-1328dt\\ \hline
     2QDressedpulse\_fixamp      &  0.0183  &   0.1837& 10&10 &  276-1328dt        \\ \hline
     2QDressedpulse\_fixang      &  0.0163  &  0.1115   &10&10 &    276-1328dt   \\ \hline
     2QDressedpulse\_fixduration &  0.0253  &  0.1146    &10&10 &   276-1328dt   \\ \hline
     2QBlockpulse  &  0.0225  &  0.1977  & 18 & 18 &  992-2496dt   \\ \hline
     2QBlockpulse\_fixamp      &  0.0161  &  0.3127  &16 &16 &  992-2496dt      \\ \hline
     2QBlockpulse\_fixang      &  0.0175 &  0.1933  &16&16   &   992-2496dt    \\ \hline
     2QBlockpulse\_fixduration &  0.0318  &  0.0619  &16&16   & 992-2496dt   \\ \hline
     Universal2QGateCircuit &  0.0253  &  0.4011  & 15 & 9   & 4768dt \\ \hline
     RXCX2QGateCircuit & 0.3971 &0.4883 &4&3& 1696dt\\ 
     \hline
  \end{tabular}
\caption{Performance metrics of various two-qubit pulse models and two-qubit gate models are provided for comparison. The table lists the expressivity (Expr), entanglement capability (Ent), number of parameters (Param), effective parameter dimension (EPD), and duration of the quantum operations (Dur) for each model. For pulse models, the amplitude, angle, and duration have been sequentially fixed and the analysis performed.}
\label{2qexpr}
\vspace{-4mm}
\end{table*}

The capacity of a circuit to generate pure states is defined as expressivity, which is introduced in previous PQC works~\cite{sim2019expressibility}. 
By taking samples from these states on the Bloch sphere, we can evaluate a quantum circuit's capacity to explore the sphere for a single qubit. This ability extends to the quantum circuit's ability to explore the Hilbert space when multiple qubits are involved.
\subsubsection{Single-Qubit Pulse}

A single-qubit pulse is a microwave pulse that is precisely tuned and controlled to manipulate the state of a single qubit in a quantum computing system. 
The effectiveness of the operation depends on the critical parameters of the pulse including amplitude, duration, and shape. 
Achieving precise and efficient manipulation of the qubit is crucial for performing quantum computations, which is why these parameters must be adjustable.
We conduct our experiments on IBM's superconducting quantum computer. Specifically, we define a single-qubit pulse whose duration and $\beta$ are obtained from the calibrated X gate from the backend and serve as its fixed parameters. The parameters are the amplitude and angle, which can fine-tune the operation on the qubit.
To explore the effects of these two parameters, we first fixed the angle to 0 and randomly sampled the amplitude 5000 times to obtain Fig.~\ref{oneqpulse} a). We then fixed the amplitude at 0.08 and randomly sampled the angle 5000 times to obtain Fig.~\ref{oneqpulse} b). From the results, we observe changes in amplitude correspond to rotations around the X-axis, while changes in angle correspond to rotations around the Z-axis. This implies that we have the potential to use a single qubit pulse with a short sequence duration to implement U3 gate operations.

\subsubsection{Expressivity of Single-Qubit Pulse and Gate}

In this section, we provide a detailed comparison between various models of single-qubit gates and our proposed model of single-qubit pulses, by defining their expressivity as a benchmark. Firstly, we consider the RZ gate, which rotates around the Z-axis when the initial state is $|0\rangle$, and all the samples on the Bloch sphere are on the $|0\rangle$ state. Secondly, we consider the Rx gate, which rotates around the X-axis when the initial state is $|0\rangle$, and the samples are distributed along a latitude line on the Bloch sphere. We then observe that the Rz+Rx circuit produces states that better cover the Bloch sphere. Next, we express the U3 gate using the Rz+Ry+Rz circuit, which can achieve rotation on the X, Y, and Z axes, resulting in higher degrees of freedom and better expressivity. 
We then demonstrate the single-qubit pulse with amplitude and angle as parameters, where amplitude indicates the amount of rotation around the X-axis and angle has enabled the rotation around the Z-axis.
We observe that the sequence duration of the single-qubit pulse is half of Rx, Rx+Rz, and Rz+Ry+Rz. 
Finally, as a reference, we uniformly sample the single-qubit matrix to simulate the most expressive circuit theoretically. Fig. \ref{oneqpulseexpr} a) shows the simulated state on the Bloch sphere after 5000 samples and the corresponding gate-level and pulse-level circuits. 

We follow the same definition of expressivity proposed in Ref. ~\cite{sim2019expressibility} for both gate-level and pulse-level ansatz 
\begin{equation}
\operatorname{Expr}=D_{\mathrm{KL}}\left(P_{\mathrm{Ansatz}}(F ; \boldsymbol{\theta}) \| P_{\text {Haar}}(F)\right),
\end{equation}
where the Kullback-Leibler (KL) divergence~\cite{kullback1951information} is a measure of distance between two probability distributions.  $P_{\mathrm{Ansatz}}(F ; \boldsymbol{\theta})$ is a distribution of state fidelities between two randomly sampled parameterized states $|\psi(\theta)\rangle$ and $|\psi(\phi)\rangle$ obtained from the ansatz
\begin{equation}
P_{\mathrm{Ansatz}}(F ; \boldsymbol{\theta})\equiv P (F = |\langle\psi(\theta) | \psi(\phi)\rangle|^2).
\end{equation}
The latter quantity $P_{\text {Haar}}(F)$ is also a state fidelity distribution but for the ensemble of Haar random states. In this case, the analytical probability density function (PDF) is known as~\cite{zyczkowski2005average}
\begin{equation}
P_{\text {Haar}}(F) = (N-1)(1-F)^{N-2},
\end{equation}
where $N$ is the dimension of the Hilbert space. If the calculated Expr is closer to $0$ for one particular ansatz, we say it is more expressive than other candidates since it is able to sample uniformly from the full Hilbert space and hence approximate any possible state. This capability is essential for variational quantum algorithms where we want to to train the ansatz to generate a particular quantum state with limited prior information about that target state.

In Fig.~\ref{oneqpulseexpr} b), the estimated confidence histograms for each circuit are displayed, with the theoretical Haar distribution shown as a reference. The bin size for generating the histograms was defined as 50, which affects the accuracy of the Kullback-Leibler (KL) divergence calculation. However, by keeping the settings consistent across all circuits, the observed results can be fairly compared. The KL divergence is reported in Table \ref{expr} to quantify the deviation, where lower KL divergence values correspond to circuit simulations that are closer to the Haar distribution, representing circuits with stronger expressive power or states closer to random states.

In Table \ref{expr}, two out of three backend models, single-qubit pulses demonstrate slightly better expressive power than single-qubit gate-level circuits. The poor performance observed in the Lima case is attributed to unfavorable backend settings. The parameter angle of a quantum pulse is equivalent to a ShiftPhase operation, which causes the qubit state to rotate around the Z-axis. We discovered that as the angle changes, the qubit state is unable to fully rotate around the Z-axis, resulting in a decrease in expressivity on the Lima backend. Generally speaking, single-qubit pulses exhibit stronger expressive power than single-qubit gate-level circuits, although it is worth noting that single-qubit pulses have only half the sequence length of Rx, Rx+Rz, and U3 circuits.

\begin{figure*}[h]
\centering
\includegraphics[width=\linewidth]{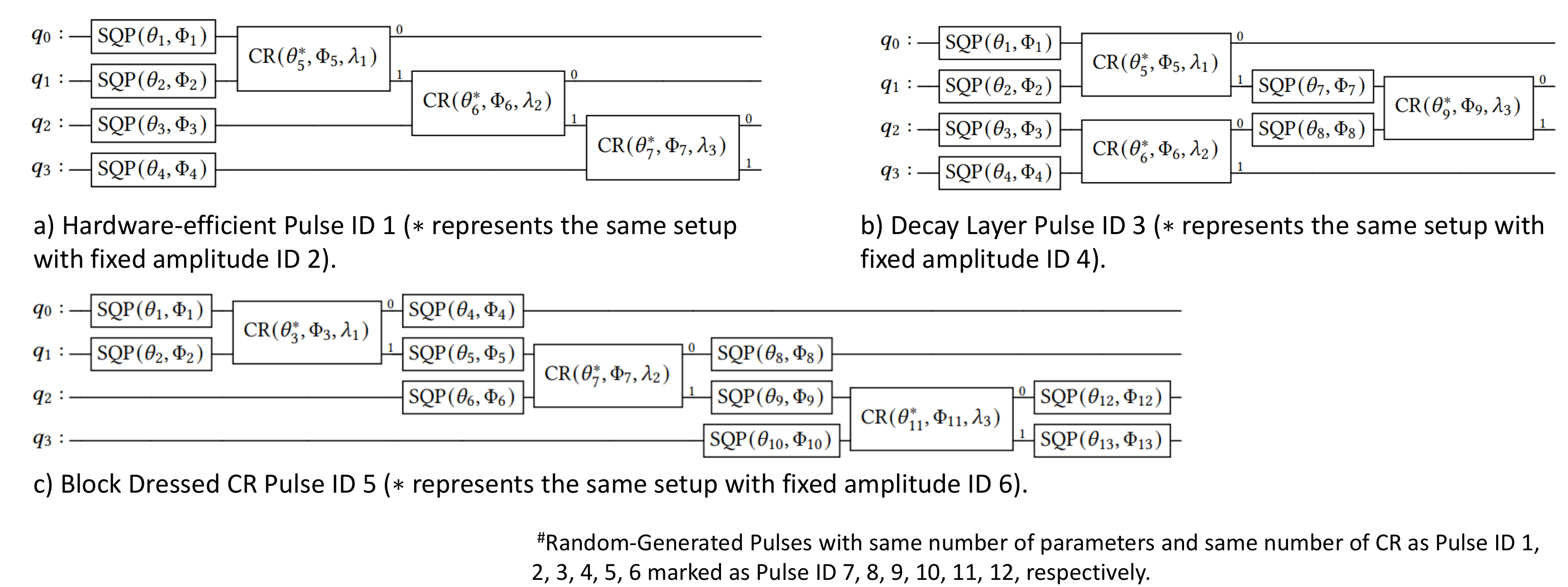}
\caption{Proposed pulse-level design space and the corresponding Pulse ID 1, 3, 5, and the proposed design space with fixed CR amplitude be marked as Pulse ID 2, 4, 6. Random-generated pulses with the same number of parameters and the same number of CR pulses as the proposed design space are introduced as Pulse ID 7, 8, 9, 10, 11, 12, i.e., Pulse ID 7 has same number of parameters and CR pulses with Pulse ID 1.}

\vspace{-2mm}
\label{policy}
\end{figure*}
\subsubsection{Multi-Qubit Pulse}
A multi-qubit pulse is a quantum operation that acts on multiple qubits simultaneously. The ability to manipulate entangled quantum states between multiple qubits is crucial in quantum computing.
The controlled-not (CNOT) gate is a common example of a multi-qubit quantum operation at the gate level, which can be implemented using multiple cross-resonance (CR) pulses and echoed pulses. 
CR is a type of multi-qubit pulse where a pulse is applied to the control qubit at the frequency of the target qubit. The cross-resonance Hamiltonian can be expressed as:
\begin{align}
            H &= Z \otimes A_2 + I \otimes B_2 \nonumber \\
            &= a_x \hat{Z} \hat{X} + a_y \hat{Z} \hat{Y} + a_z \hat{Z} \hat{Z} + b_x \hat{I} \hat{X} + b_y \hat{I} \hat{Y} + b_z \hat{I} \hat{Z}.
\end{align}
At the gate level, cross-resonance pulses need to be elaborately calibrated to implement desired function. Specifically, we want to keep the $ZX$ term and eliminate other interactions. Such calibration process requires techniques including echoed CR and phase calibration as discussed in~\cite{sheldon2016procedure}. At the pulse level, however, there is no need to remove terms other than $ZX$. In this way, we trade precise controllability for more flexibility and capacity to further explore the Hilbert space.

\begin{figure}[t]
\centering
\includegraphics[width=\linewidth]{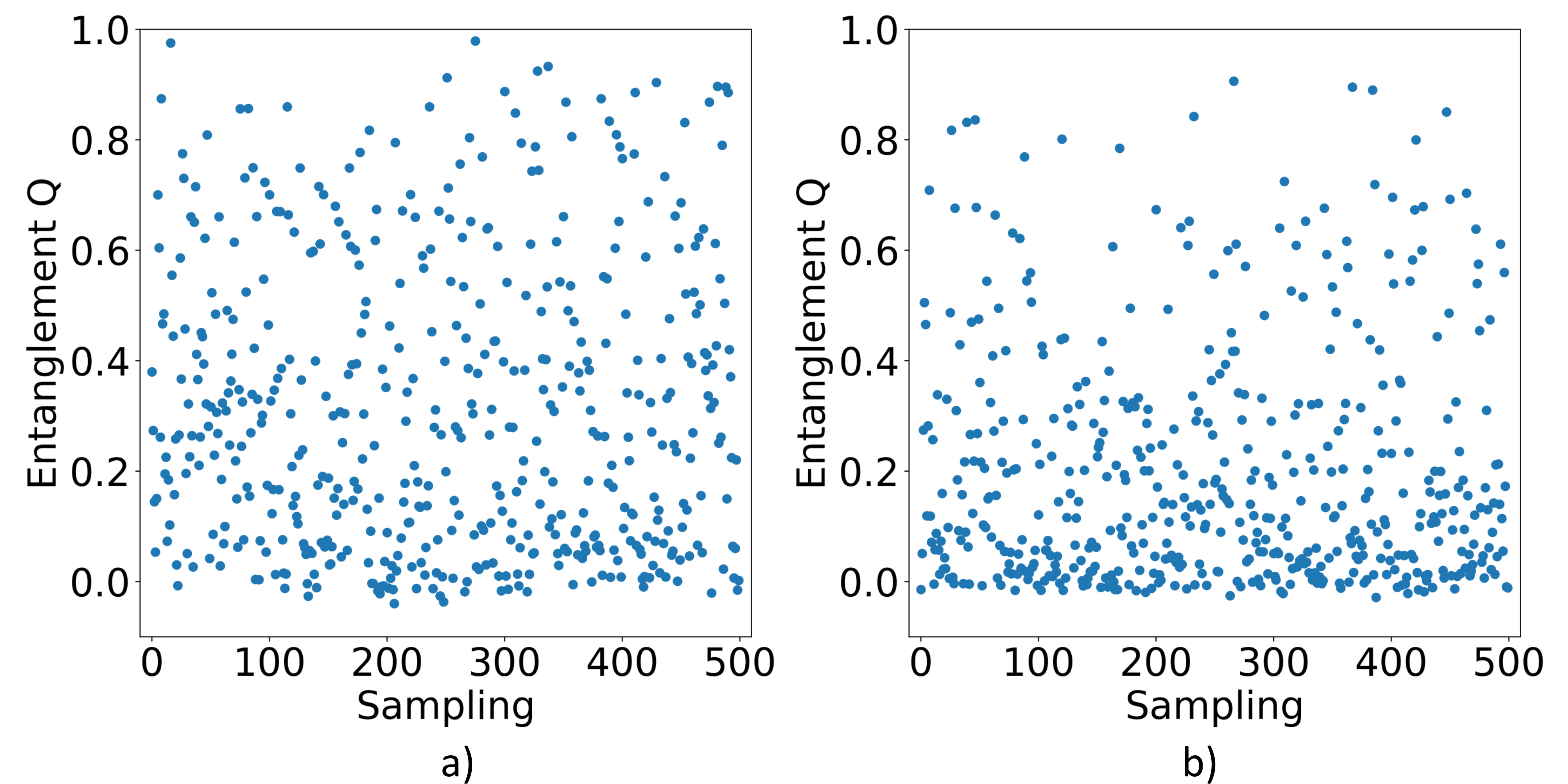}
\caption{Q measurement result through 500 times sampling of 2QBlockpulse and 2QBlockpulse\_fixamp.}

\vspace{-6mm}
\label{entmap}
\end{figure}

\subsection{Entanglement Capability}
Entanglement is a unique quantum phenomenon that refers to the non-classical interdependence between two or more qubits, even when they are located far apart. In variational quantum algorithms and quantum machine learning, generating shallow circuits with strong entanglement capabilities is essential. Consequently, understanding how to measure a quantum circuit's entanglement ability is crucial in selecting an appropriate design space. This helps to create more effective solution spaces for tasks such as ground state preparation or machine learning benchmarks while also preserving the non-trivial correlations in quantum data.

One commonly used standard for measuring the entanglement ability of a quantum circuit is the Meyer-Wallach (MW) entanglement $Q$-measure. It can be employed to quantify the global entanglement measure of pure-state qubits entanglement, with the benefits of scalability and ease of computation.
We define the entanglement capability as the average $Q$-measure from an ensemble of randomly sampled states
\begin{equation}
\text { Ent }=\frac{1}{|S|} \sum_{\boldsymbol{\theta}_i \in S} Q\left(\left|\psi_{\boldsymbol{\theta}_i}\right\rangle\right),
\end{equation}
where the Meyer-Wallach $Q$-measure is proposed to estimate the number and types of entangled states an ansatz can generate. It can be calculated though the average of the
purity of each qubit~\cite{brennen2003observable}
\begin{equation}
Q(|\psi\rangle)=2\left(1-\frac{1}{n} \sum_{k=1}^n \operatorname{Tr}\left[\rho_k^2\right]\right),
\end{equation}
where $\rho_k$ is the reduced density matrix of the $k$-th qubit. Notice for any product state $|\psi\rangle \otimes |\phi\rangle \otimes \cdots$, $Q=0$, while $Q=1$ for the Greenberger-Horne-Zeilinger (GHZ) state $|\psi\rangle_{\text{GHZ}} = (|0\rangle^{\otimes N} + |1\rangle^{\otimes N})/\sqrt{2}$.
\begin{figure*}[t]
\centering
\includegraphics[width=\linewidth]{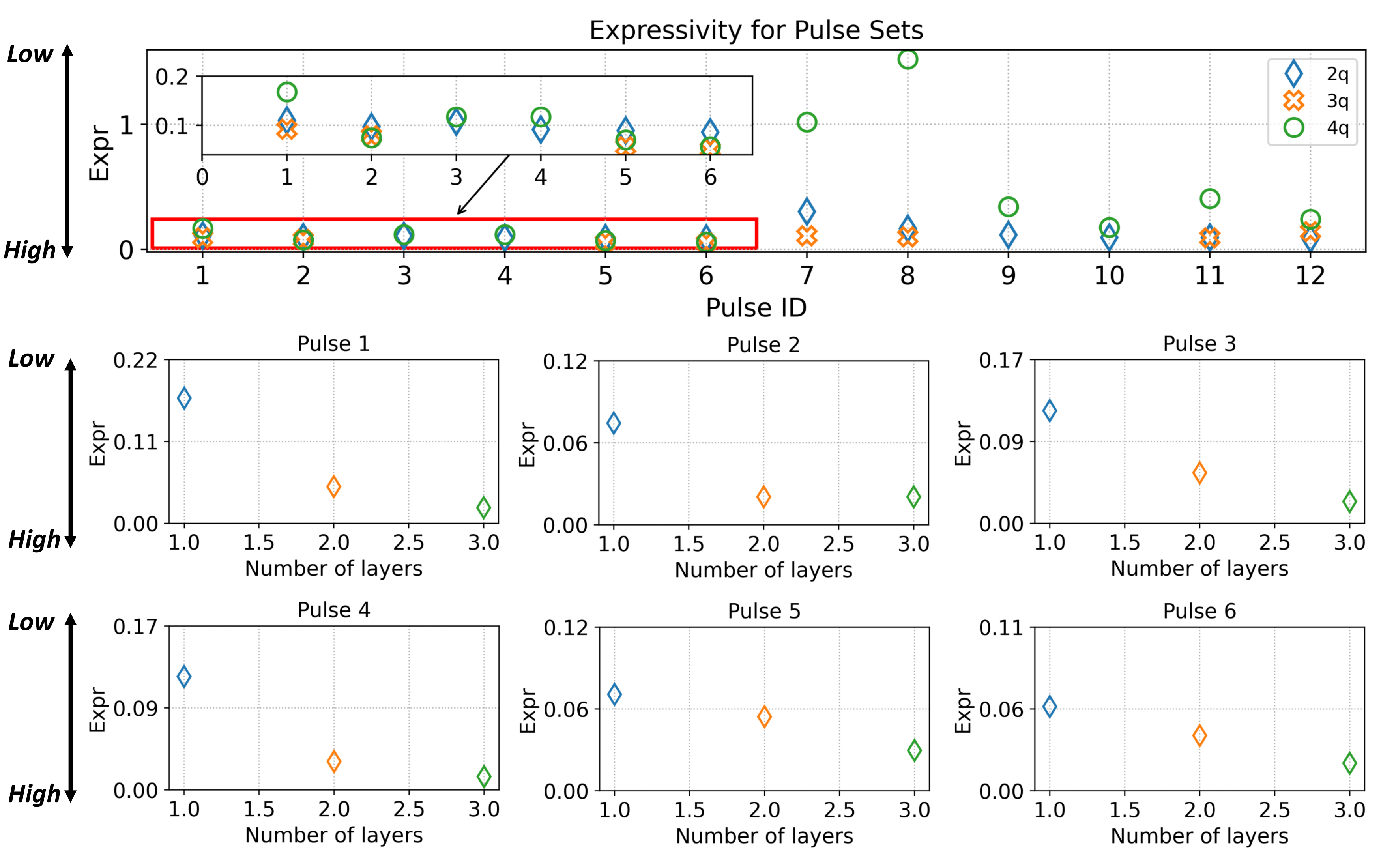}
\caption{Illustration of the change in the expressivity of various pulse sets as the number of qubits increases. The main plot shows the overall trend, with a zoomed-in section highlighting the results of six proposed pulse-level design spaces. The figure demonstrates that expressivity increases with an increasing number of qubits until a small decrease in expressivity is observed when the number of qubits reaches four, which may be due to insufficient circuit depth. Additionally, the six subplots show the change in the expressivity of each of the pulse-level design spaces as the number of layers increases for four qubits.}

\vspace{-2mm}
\label{exprtemplate}
\end{figure*}
In order to explore the entanglement capabilities of parameterized pulse circuits, we constructed two types of two-qubit parameterized pulse circuits as shown in the Fig.~\ref{2qpulse}. We then sampled and calculated the entanglement capabilities of the corresponding circuits. In particular, we use the Hamiltonian tomography of the CR to investigate the impact of phase and duration variations on the components of IX, IZ, and IY, and result in altering the proportion of the ZX term, which directly affects the entanglement power \cite{sheldon2016procedure, alexander2020qiskit}. Thus, it is necessary to analyze each parameter's effect in CR. As a result, we measured the entanglement and expressivity capabilities while fixing the amplitude, angle, and duration, respectively, to observe the effects of the corresponding parameters. 

In Table~\ref{2qexpr}, we introduce two pulse models within the structure as shown in Fig.~\ref{2qpulse}, and fix each type of parameter sequentially. We also put two gate models as baseline to compare, one is the Universal2QgateCircuit, a quantum circuit composed of 18 gates from CNOT, Ry, and Rz being the minimal gate-based circuit that simulates an arbitrary two-qubit unitary operator up to global phase~\cite{shende2004minimal}. Another gate model consists of four Rx gates and one CNOT gate, where two of Rx gates are in front of the CNOT implemented on the first and second qubit, respectively, and another two Rx gates are behind the CNOT. From Table \ref{2qexpr}, fixing the amplitude (2QDressedpulse\_fixamp and 2QBlockpulse\_fixamp) resulted in a significant improvement in the entanglement capability and a slight improvement in expressivity, while fixing the angle (2QDressedpulse\_fixang and 2QBlockpulse\_fixang) resulted in a small improvement in expressivity but did not significantly affect the entanglement capability. Fixing the duration (2QDressedpulse\_fixduration and 2QBlockpulse\_fixduration) resulted in a negative impact in all metrics. Overall, Universal2QGateCircuit had the strongest entanglement capability, while 2QBlockpulse\_fixamp had the best expressivity. Pulse models generally exhibit better expressivity and have shorter sequence durations compared to gate models. Although the pulse-level circuit exhibits a relatively low entanglement capability than the gate-level circuit, it still possesses points with Q values approaching 1 as shown in the Fig. \ref{entmap}. This characteristic is advantageous for variational quantum algorithms (VQAs) tasks, as it allows us to search for and identify points with Q values close to 1 during the training and optimization processes.

\subsection{Effective Parameter Dimension}
\begin{figure*}[t]
\centering
\vspace{-12mm}

\includegraphics[width=\linewidth]{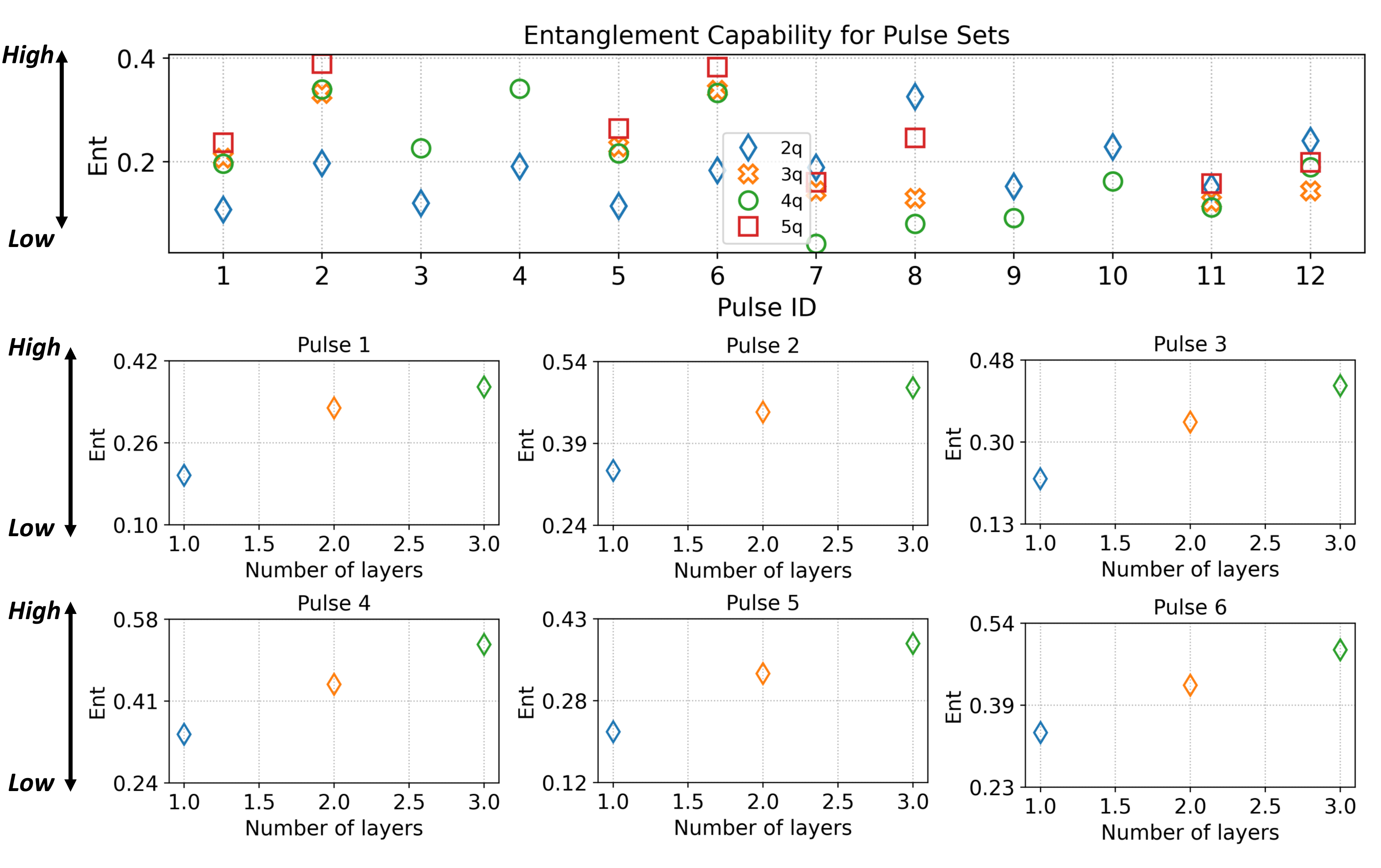}
\caption{The main plot of this figure shows the entanglement capability of various pulse sets as the number of qubits increases. The zoomed-in section of the plot displays the results of six pulse-level design spaces proposed in this study. The plot illustrates that as the number of qubits increases, the entanglement capability continues to improve. The six subplots depict the entanglement capability of each of the six pulse-level design spaces as the number of layers increases, with the qubit count fixed at four. The subplots demonstrate that with an increasing number of layers, the entanglement capability of each pulse-level design space also continues to increase.}

\vspace{-2mm}
\label{enttemplate}
\end{figure*}

Some parameters of PQC can be eliminated without affecting its expressivity. 
Effective parameter dimension (EPD) is a measure that quantifies the number of independent parameters of quantum states generated by PQC~\cite{PRXQuantum.2.040309}. It reflects how many independent directions the parameters can explore over the Hilbert space. 

EPD can be calculated through Quantum Fisher information (QFI)~\cite{Liu_2020}, which characterizes the sensibility of the state generated by PQC to its parameter changes~\cite{RevModPhys.94.015004}. Concretely, the QFI matrix is
\begin{equation}
\hspace{-2mm}\mathcal{F}_{i j}=\operatorname{Re}\left(\left\langle\partial_i \psi(\boldsymbol{\theta}) \!\mid\! \partial_j \psi(\boldsymbol{\theta})\right\rangle-\left\langle\partial_i \psi(\boldsymbol{\theta}) \!\mid\! \psi(\boldsymbol{\theta})\right\rangle\left\langle\psi(\boldsymbol{\theta}) \!\mid\! \partial_j \psi(\boldsymbol{\theta})\right\rangle\right)
\end{equation}
Here the gradient of parameters is calculated by finite difference: 
\begin{equation}
\partial_i \psi(\boldsymbol{\theta}) \approx\left[\psi\left(\boldsymbol{\theta}+\epsilon \mathbf{e}_i\right)-\psi\left(\boldsymbol{\theta}-\epsilon \mathbf{e}_i\right)\right] / 2 \epsilon.
\end{equation}
EPD is the rank of QFI matrix: $d_{\mathrm{eff}}=\operatorname{rank} \mathcal{F}(\boldsymbol{\theta})$,
i.e., the number of linearly independent parameters over the design space.

The experiments presented in Table~\ref{2qexpr} suggests that all parameters of the pulse-level model proposed in the Fig.~\ref{2qpulse} are valid. Compared to two-qubit gate models, the parameter dimensions at the gate level are more susceptible to redundancy, with Universal2QGateCircuit exhibiting six redundant parameters.
Based on our analysis in the previous section, we conclude that the richness of EPDs is a significant contributing factor to the superior expressivity of the PPC model over gate-based PQC within a shorter duration.

\begin{table*}[t]
\centering

  \captionsetup{}
  \setlength{\tabcolsep}{2.2em}
\begin{tabular}{|c|c|c|c|c|}
\hline
\textbf{Pulse ID} & \textbf{\# of Params} & \textbf{EPD} & \textbf{\# of CR Pulses} & \textbf{Pulse Circuit Depth} \\ \hline
1                 & (5N-3)L               & (5N-3)L      & (N-1)L                   & NL                           \\ \hline
2                 & 2(2N-1)L              & 2(2N-1)L     & (N-1)L                   & NL                    \\ \hline
3                 &   (6N-3)L                    &       (6N-3)L       &    (N-1)L                      &    4L                          \\ \hline
4                 & (5N-2)L               &  (5N-2)L               &    (N-1)L                      & 4L                             \\ \hline
5                 & (9N-7)L               & (9N-7)L      & (N-1)L                   & (2N-1)L                      \\ \hline
6                 & (8N-6)L               & (8N-6)L      & (N-1)L                   & (2N-1)L                      \\ \hline

\end{tabular}
\caption{The estimated costs associated with the proposed design space at the pulse level, which includes the number of parameters, effective parameter dimension, two-qubit operations, and circuit depth, are expressed in terms of two variables: N (number of qubits) and L (number of circuits).}
\label{pulseparameter}
\vspace{-4mm}
\end{table*}
\section{Proposed Pulse-level Design Space}
\subsection{Insights of Proposed Design Space}
In this section, we introduce three different PPC model designs with a polynomial scaling of parameters. The first PPC template is the hardware-efficient pulse in Fig.~\ref{policy} a, where we only apply two-qubit CR pulse on physically connected qubits. Hence, such a pulse template is efficient in the sense that requires least compiling resources to implement on the hardware. The second pulse template is a decay layer pulse in Fig.~\ref{policy} b. designed to improve the trainability of PPC. The hope is to gradually reduce the operating scope (involves fewer qubits) and hence alleviate the barren plateaus (BP) phenomenon~\cite{mcclean2018barren}. Most studies on BP are conducted on the gate-level, and similar research on the pulse level is minimally explored.
The last pulse template is the block-dressed CR pulse in  Fig.~\ref{policy} c. We first dress the native two-qubit CR pulse with four single-qubit pulses on both sides to enhance expressivity. Then, we arrange such a dressed CR pulse as a block and loop through all the qubits. In general, one can always adopt neural architecture search (NAS) methods to construct an automatic workflow to build customized PPC models. But those approaches are usually too expensive to run on real quantum hardware. The proposed pulse templates in this study are aiming to make pulse-level research more accessible to the community.

\subsection{Evaluation and Analysis of Design Space}
In this section, we conduct an analysis of the proposed pulse-level design space and summarize our observations and insights for each analyzed criterion. These analyses are performed with respect to all evaluation criteria and highlighted general trends. Before introducing our observations and insights, we first conduct a detailed investigation using the criteria mentioned in the previous section, in order to better understand the insights obtained from the subsequent presentation of the comprehensive trends. We include evaluations of expressivity, entanglement capability, as well as cost estimation for the proposed pulse-level design space.

\subsubsection{Expressivity Analysis.} Firstly, we measure the ability and potential of the proposed pulse-level design space to explore the Hilbert space, while adding random-generated pulses without any design space-based policy as a comparison to the experiments.
By evaluating the results of a series of expressivity tests (see Fig.~\ref{exprtemplate}), the value of K-L divergence drops gradually as the test depth (the number of qubits) increases, indicating that the expressivity of the pulse-level design space increases progressively. As the number of qubits exceeds 4, however, the value of K-L divergence increases slightly, indicating that one layer of pulse-level design space is incapable of exploring this portion of the Hilbert space. This is observed for all the proposed pulse-level design spaces we described. In contrast, the trend of expressivity with the number of qubits for pulses generated at random without policy is erratic and unpredictable.
\begin{figure*}[t]
\centering
\includegraphics[width=\linewidth]{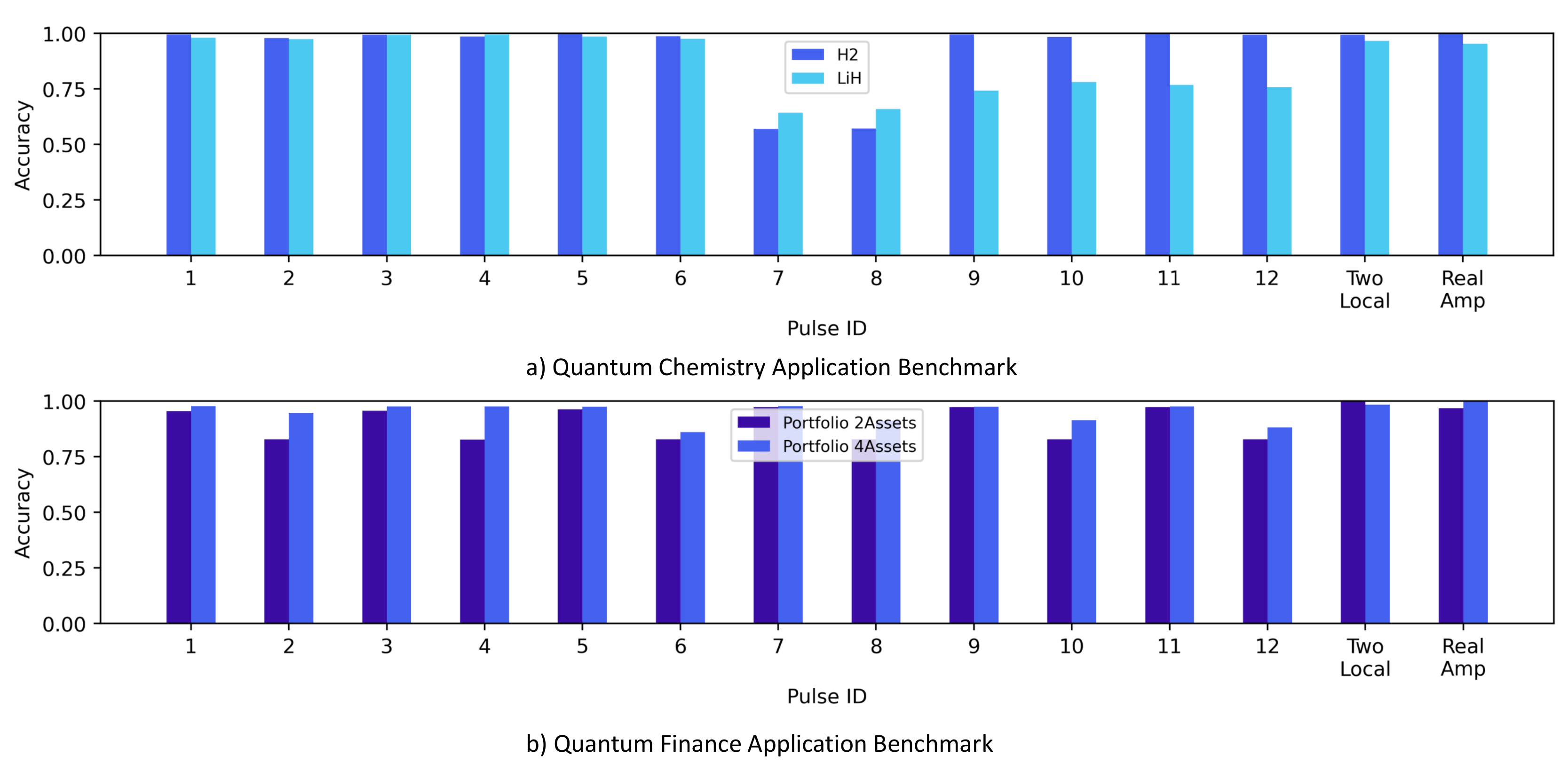}
\caption{The application benchmark of the proposed pulse-level design space, compared to random-generated PQC and two gate-level PQCs that are verified to have good performance. a) displays the results for quantum chemistry applications, with tasks involving the electronic structure of $H_2$ and $LiH$ molecules, the bond lengths default to 0.735\AA ~and 1.5\AA, respectively. b) displays the results for quantum finance applications, with tasks involving portfolio optimization for 2 and 4 assets.}
\vspace{-2mm}
\label{applicationbenchmark}
\end{figure*}

Then, we fix the number of qubits at four and gradually increase the width of the test, that is, for the number of layers of PPC. In Fig.~\ref{exprtemplate}, for each of the proposed pulse-level design spaces, increasing the number of circuit layers significantly improves expressivity. In addition, we find that the design space with a fixed CR amplitude is significantly more expressive than the corresponding design space without a fixed CR amplitude, and that among all design spaces, the design space with Pulse ID 2, which is the design space Dressed CR with a fixed CR amplitude, obtains the best performance in expressivity for a circuit with 1 layers and two, three, or four qubits, respectively. When there are two circuit layers, the hardware-efficient pulse with a Pulse ID 6 that fixes the amplitude of the CR gets the highest expressivity performance. When there are three circuit layers, the Decay-layer pulse with Pulse ID 4 that fixes the amplitude of CR has the best expressivity performance.

\subsubsection{Entanglement Capability Analysis. }Secondly, we tested the entanglement capability of the suggested pulse-level design space to investigate and assess the strength of the proposed pulse-level design space's entanglement capability. Likewise, no design space-based policy is added to the experiment for comparison purposes. According to Fig.~\ref{enttemplate}, the entanglement capability of all pulse-level design spaces grows progressively as the test depth (the number of qubits) rises. Meanwhile, entanglement capability with respect to the number of qubits for pulses generated at random without a policy is unpredictable. Then, we fix the number of qubits at four and increase the test's width progressively. From the six subplots, we can conclude that the progressive growth of the layer of PPC has a higher impact on the entanglement capability than the rise in qubits. Experiments on entanglement capability revealed that the design space with a fixed CR amplitude produces greater entanglement capability than the design space without a fixed CR amplitude. In the case of a single-layer circuit, the hardware-efficient pulse fixed CR amplitude with Pulse ID 2 can create the strongest entanglement capability. In the design space without fixed CR amplitude, the Decay-layer pulse with Pulse ID 3 provides the highest entanglement capability. When the number of circuit layers increases to two and three, the Decay-layer pulse with fixed CR amplitude and a Pulse ID 4 provides the best entanglement performance.
\begin{table*}[t]
\centering

  \captionsetup{}
  \setlength{\tabcolsep}{1.1em}
\begin{tabular}{|c|c|c|c|c|c|}
\hline
\textbf{Applications}              & \textbf{Benchmarks} & \textbf{\# of Qubits} & \textbf{\# of Pauli Strings} & \textbf{Measurement Bases} & \textbf{Entanglement Q} \\ \hline
\multirow{2}{*}{Quantum Chemistry} & $H_{2}$        &2        & 5                            & X, Y, Z                    & 0.04914                 \\ \cline{2-6} 
                                   & $LiH$          &4       & 100                          & X, Y, Z                    & 0.00072                 \\ \hline
\multirow{2}{*}{Quantum Finance}   & 2 Assets      &2     & 3                            & Z                          & 0                       \\ \cline{2-6} 
                                   & 4 Assets      &4     & 10                           & Z                          & 0.00097                 \\ \hline
\end{tabular}
\caption{A detailed characterization of several application benchmarks using the number of qubits, number of Pauli strings, measurement basis, and entangled qubits required to find the ground state of the problem. }
\label{benchmarktable}
\vspace{-4mm}
\end{table*}

\subsubsection{Estimated Cost Analysis}
For each given pulse-level design space, we estimate the cost of implementation in Table~\ref{pulseparameter} in terms of the number of qubits N and the number of layers L of the circuit. We assess the cost of implementation for each pulse-level design space, taking into account the number of parameters, effective parameter dimension, number of CR pulses, and pulse width. We see a linear relationship between the number of parameters, effective parameter dimension, and number of CR pulses for all design spaces and the number of qubits. In addition, the pulsed circuit depth of all design spaces is proportional to the number of qubits.

\section{Application Benchmarking}
\subsection{Benchmark Applications}
\subsubsection{Quantum Chemistry.}
Quantum chemistry studies the electronic structure and reaction characteristics of molecules using the ideas and tools of quantum mechanics to describe and calculate these properties. Quantum chemistry seeks to answer the following challenge in particular: how to determine the ground state energy, electronic structure, and features such as energy barriers for a given molecule. Using the proposed pulse-level design space, we intend to achieve the variational quantum eigensolver (VQE) to solve the challenge of determining the ground state energy of molecules in quantum chemistry as part of our application benchmarking. This process can be considered as the solution of an eigenvalue problem for a Hamiltonian-like matrix, where the Hamiltonian describes the energy of particle interactions in a molecule and the eigenvalue is the ground state energy. The number of qubits, the number of Pauli strings, the measurement base, and the theoretical requirements of the ground state of the task for entanglement are stated for both molecular problems in Table~\ref{benchmarktable}.
\subsubsection{Quantum Finance.}
Quantum finance refers to the multidisciplinary application of quantum computing technology to the financial sector. Using quantum computers, it intends to handle a variety of financial problems, including but not limited to risk management, portfolio optimization, pricing of financial derivatives, etc.
For our application benchmarking, we have selected the problem of investment optimization. Investment optimization is the construction of the optimal portfolio to achieve a particular purpose, given an underlying investment and a set of constraints. The risk factor is set to 0.5, and we are provided the two assets and four assets, as well as the randomly produced `historical data' for each, and apply information to form the quadratic problem, then we map the variables in the quadratic problem to qubits so that translate the problem to Ising Hamiltonian. Our objective is to achieve VQE using pulse-level design spaces in order to obtain the ideal portfolio model. The information about the quantum finance problems is also provided in Table~\ref{benchmarktable}.

\subsection{Parameters Constraint Generator}
\begin{figure}[t]
\centering
\includegraphics[width=0.9\linewidth]{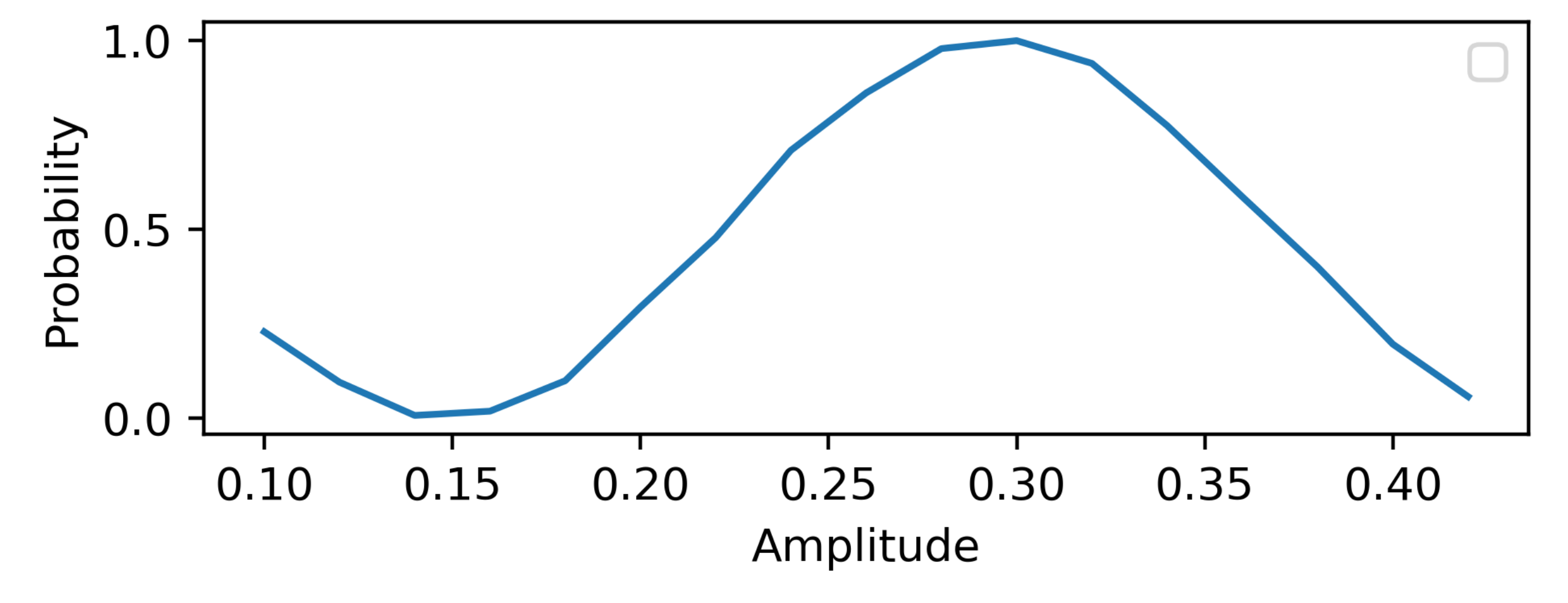}
\caption{
Dynamics of quantum states of an SQP with amplitude in the range of [0.1, 0.4] on $ibmq\_guadalupe$.}
\vspace{-5mm}
\label{amp}
\end{figure}

Parameterized pulse circuits operate by specifying a sequence of pulses that must adhere to hardware-imposed limitations. First, the pulse amplitude must be expressed as a fraction of the maximum output voltage of the arbitrary waveform generator (AWG) and must therefore be constrained to the interval [-1,1] \cite{egger2023study}. Further experimentation by us has shown that the amplitude does not need to occupy the entire [-1,1] interval to achieve the desired oscillation period on the quantum state, and that the range may differ slightly across different hardware backends. The variance between the overall minimum and maximum values is typically around 0.35. For example, in Fig. \ref{amp}, the amplitude range is [0.1, 0.4] on $ibmq\_guadalupe$, thus, the training cost could be greatly reduced by selecting a suitable range of amplitude instead of [-1,1]. Second, the pulse angle affects the orientation of the rotation axis in the plane and thus oscillates within the interval [0,2$\pi$]. Third, the duration of a pulse, expressed in the unit of dt, must be a multiple of 16 to be loaded in the AWG, and the duration must be kept long enough to prevent the width of the top except for the rise-fall edge from being negative. To impose these constraints, we have introduced a parameter constraint generator to restrict the distribution range of all the parameters. For amplitude, we benchmarked multiple backends and provided a lookup-table (LUT) to find the amplitude range corresponding to a particular backend. For angle, we set this parameter to be within the range [0,2$\pi$]. For the duration, we restrict this parameter to the range [256,1024].

\subsection{Evaluation and Analysis}
All the application benchmarks were conducted on Qiskit-Dynamics and Qiskit-Aer. Fig.~\ref{applicationbenchmark} depicts a comparison of task performance between the suggested pulse-level design space, pulses generated at random, and the gate-based approach. We chose TwoLocal, which consists of Ry gates and CZ gates, and RealAmplitude, which consists of Ry gates and CX gates, for the gate-based model. There is no significant difference between the proposed design space and the randomly produced pulses for quantum finance tasks. We are able to identify the design space with a fixed CR amplitude that performs better in terms of expressivity and entanglement capability but yields poorer application results. This phenomenon is due to the factor we discussed in the previous section that our tested duration is mapped discretely to unit `dt' from hardware to software, thus, the power of parameter `duration' is limited.
Thus, the criteria could be effective for guiding circuit design the majority of the time, but due to the fact that the proposed pulse-level design space must be evaluated further, application benchmarking is essential. In addition, the gate-based model generally obtains better performance than the pulse-based model in these tasks.

As we evaluate the quantum chemistry tasks, we find in the $H_2$ task, the random-generated pulse with few numbers of parameters (pulse ID 6 and 7, within the same number of parameters of Hardware-efficient pulse) performs badly. As for the $LiH$ task, the proposed pulse-level design space (Pulse ID from 1 - 6) achieves way better results than random-generated pulses, and also slightly better than the gate-based model. The reference ground state energy is -8.9407H, and the estimated energy of Pulse ID 1, 3, and 5 are -8.7766H, -8.8946H, and -8.8114H, respectively, the `TwoLocal' gate-based ansatz obtains -8.6442H, as for `RealAmplitude' gate-based ansatz, the result is -8.5214H. In terms of the sequence duration of the circuit, we have 979dt, 1835dt, and 1954dt for Pulse ID 1, 3, and 5, whereas, `TwoLocal' and `RealAmplitude' with 24800dt and 5600dt, respectively. In conclusion, the proposed pulse-level design space performs up to 4.2\% in terms of the accuracy of the selected quantum chemistry problem with 67.2\% latency advantage in duration, and up to 96\% shortening in terms of duration with 1.5\% accuracy advantage compared to the selected gate-based design.

The different observations from the quantum finance task and the quantum chemistry task are illustrated in Table~\ref{benchmarktable}. The portfolio optimization problem with 2 assets and 4 assets are both too simple for VQEs, as they have a light number of Pauli strings and nearly no entanglement requirement by the ground state of the problem, and the measurement base is only on Z-axis. Thus, these problems can be even solved by random-generated pulses with good performance and lead to tiny differences that are hard to compare between models. In contrast, quantum chemistry problems have much more Pauli strings, stronger entanglement requirements, and more complex measurement bases than on all of X, Y, and Z, and thus, we observe the expectations in the quantum chemistry benchmark.

\section{discussion}
The ability to manipulate quantum states of qubits at the pulse level holds great promise, as it provides fine-grained control over the qubits. Parameterized quantum pulse enables researchers to explore new design space and generate more customized and complex quantum operations by leveraging the specific physical properties of the quantum processors. For instance, we have shown that a parameterized pulse template can generate highly entangled states with a high degree of expressivity, indicating the potential of designing parameterized pulse as the building block for the ansatz of variational quantum algorithms. Furthermore, the pulse naturally allows for flexible tuning with sufficient parameters, allowing for the exploration of Hilbert space fully~\cite{5439324}. In terms of EPD, the high EPD value demonstrated in our experiments also indicates the low redundancy of parameterized quantum pulse, ensuring efficient utilization of experimental resources.

However, pulse-level control also has its limitations. One challenge is the complexity of designing pulse sequences for arbitrary quantum operations, as pulse-level control heavily relies on the physical implementation of the qubits. Consequently, designing parameterized quantum pulse requires a deep understanding of the underlying interactions and may require significant computational resources. Another challenge is the impact of noise and decoherence on the system. Basis gates can be fine-tuned for parameterized gates, but the calibration of arbitrary pulses is generally difficult. These effects can limit the effectiveness of pulse-level control, especially when dealing with large-scale quantum systems. The formalism of quantum pulses is still an incomplete field, making it challenging to design quantum protocols at the pulse level.
\begin{figure}[t]
\centering
\includegraphics[width=0.9\linewidth]{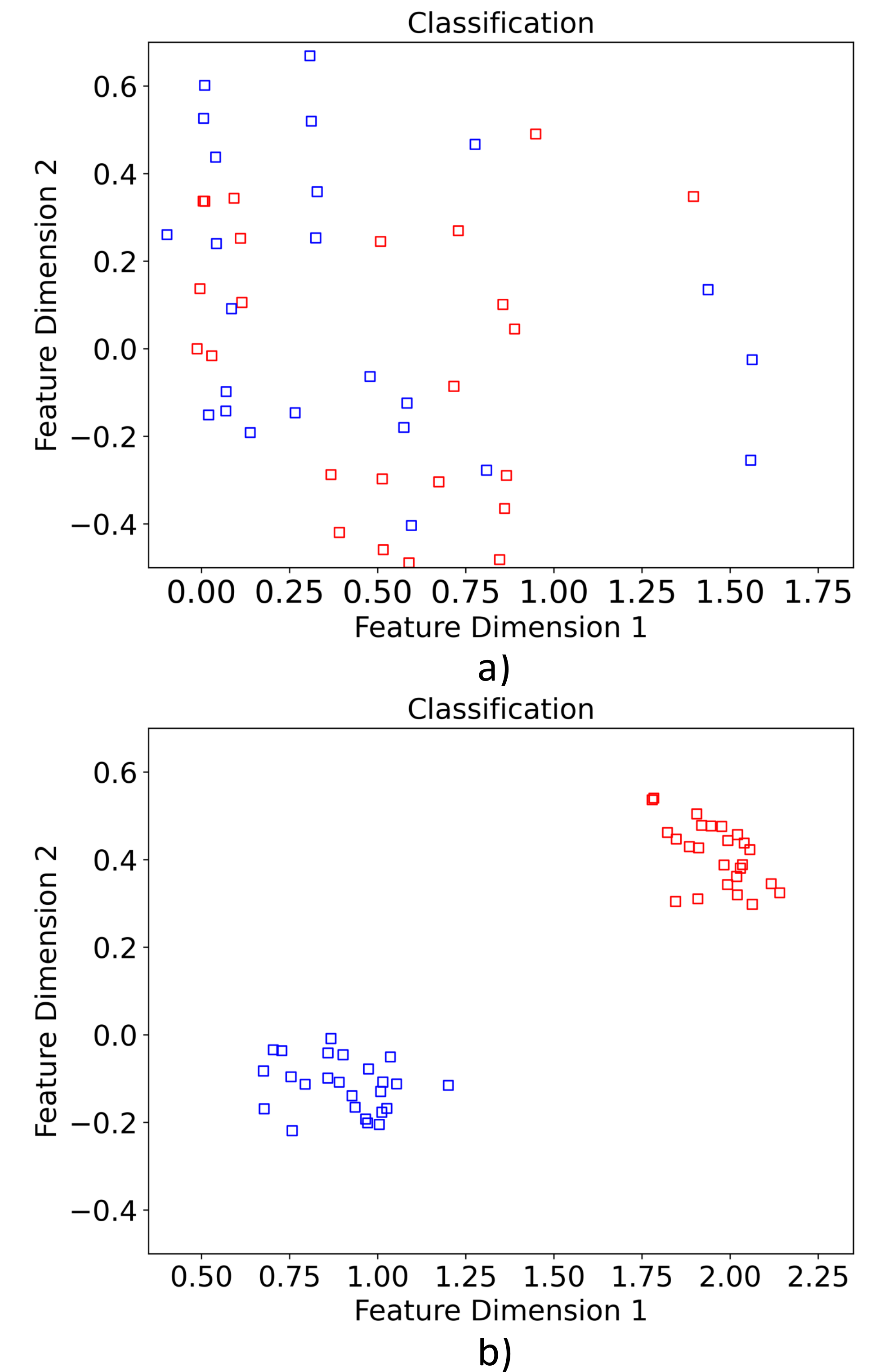}
\caption{
Binary classification of random-generated datasets with 16 features. a) Encoding by 8 SQPs on two qubits. b) Encoding by 16 single-qubit rotation gates on two qubits.}
\vspace{-4mm}
\label{encode}
\end{figure}

Moreover, we have encoded classical data using a parameterized pulse encoder, which we initially believed had the potential to encode more data in a shorter amount of time. However, as shown in Fig.~\ref{encode}, in a simple experiment, we found that the gate-based encoder produced superior results, even with a shallow circuit. Although encoding classical data with parameterized pulses remains intriguing, this observation is unfavorable, and further research is necessary. We are still possible gain some benefits from a parameterized pulse encoder, for example, it is possible to encode time-sequence datasets by encoding time information into the duration parameter in parameterized pulse. 

To overcome the limitations of quantum pulse applications, researchers are exploring new approaches that combine pulse-level and gate-level control. For example, in this paper, we designed multiple pulse templates from gate-inspired ansatz with good performance. Such an approach allows for the flexibility and fine-grained control of pulse-level control while maintaining the simplicity and generality of gate-level protocols. Another approach is to use machine learning algorithms to optimize the pulse sequences for specific quantum tasks. This approach can reduce the complexity of programming the hardware and improve the overall performance of the quantum computation~\cite{PhysRevA.103.022613}.

Overall, the advantages and limitations of parameterized quantum pulses highlight the need for continued research and development in this area to fully realize the potential of quantum computing. In this paper, we have evaluated parameterized quantum pulses from various perspectives, including expressivity, entanglement capability, effective parameter dimension, duration, and application-oriented benchmarking, which we hope will pave the way for future studies of quantum pulses.

\clearpage
\bibliographystyle{ACM-Reference-Format}
\bibliography{sample-base}

\appendix









 \section{Acknowledgement}
\label{sec:acknowledgement}
We would like to express our gratitude for the utilization of IBM Quantum services and PaddleQuantum in conducting this research. Additionally, our appreciation extends to Dr. Yao Lu for the insightful discussions of this work.
\end{document}